\documentclass[journal=jcisd8,manuscript=article]{achemso}

\usepackage[version=3]{mhchem} 
\usepackage{hyperref}
\usepackage{caption}
\usepackage{subcaption}
\usepackage{amsfonts}
\usepackage{textcomp}
\usepackage{gensymb}
\usepackage{moresize}
\usepackage[dvipsnames]{xcolor}
\author{Ian Slagle}
\affiliation{Material Science and Engineering, Georgia Institute of Technology, Atlanta, GA, USA}
\author{Faisal Alamgir}
\affiliation{Material Science and Engineering, Georgia Institute of Technology, Atlanta, GA, USA}
\author{Victor Fung}
\affiliation{Computational Science and Engineering, Georgia Institute of Technology, Atlanta, GA, USA}
\email{victorfung@gatech.edu}

\title[]{Determining Atomic Structure from Spectroscopy via an Active Learning Framework}

\abbreviations{IR,NMR,UV}
\keywords{American Chemical Society, \LaTeX}

\begin{document}







\begin{abstract}
  Determining atomic structure from spectroscopic data is central to materials science but remains restricted to a limited set of techniques and material classes, largely due to the computational cost and complexity of structural refinement. Here we introduce ActiveStructOpt, a general framework that integrates graph neural network surrogate models with active learning to efficiently determine candidate structures that reproduce target spectra with minimal computational expenditure. Benchmarking with X-ray pair distribution function data, and with the more computationally demanding simulations of X-ray absorption near-edge spectra (XANES) and extended X-ray absorption fine structure (EXAFS), demonstrate that ActiveStructOpt reliably determines structures that match closely in spectra across diverse materials classes. Under equivalent computational budgets, ActiveStructOpt outperforms existing structure determination methods. By enabling data-efficient, multi-objective structural refinement across a broad range of computable spectroscopic techniques, ActiveStructOpt provides a flexible and extensible approach to atomic structure determination in complex materials.  
\end{abstract}


\section{Introduction}

Effectively characterizing materials relies on using spectroscopy to construct a model of the atomic structure. Determining such a structure model involves iteratively suggesting candidate structures and simulating the corresponding spectra until a structure is proposed with a simulated spectrum that agrees with experiments. Different spectroscopies are sensitive to different aspects of the structure, and for complex materials, uniquely determining the structure often requires multiple spectroscopies. However, the computational expense of simulating many types of spectroscopy makes the iterative process within existing solvers intractable. Our proposed new solver actively trains a graph neural network (GNN) to efficiently leverage simulations to determine the structure from one or more spectroscopies. 

The most common experimental spectroscopy for determining the atomic structure is diffraction \cite{runcevski2021rietveld, bergerhoff1983inorganic}. Structural determination from diffraction often uses one of two classes of solvers for both Rietveld and pair distribution function (PDF) refinements. Gradient-based solvers, such as Levenberg-Marquardt \cite{levenberg1944method}, use analytical or estimated gradients to determine the next iterative structure candidate \cite{mittemeijer2013modern, rietveld1969profile, farrow2007pdffit2}. Without analytical gradients, these methods require simulating many structures to estimate gradients. Additionally, as local techniques, these solvers struggle when applied to global structural determination. By contrast, probabilistic solvers, such as simulated annealing \cite{mittemeijer2013modern} and reverse Monte Carlo \cite{mcgreevy2001reverse, billinge2008local} more effectively solve global optimization problems and more efficiently search high-dimensional spaces (in the absence of analytical gradients) by evaluating steps via a Metropolis-Hastings condition rather than selecting a step by the gradient \cite{mcgreevy2001reverse}. Probabilistic techniques can require simulating several thousands of candidate structures \cite{timoshenko2014analysis}. Despite requiring either analytical gradients or many simulations, both classes of solvers remain computationally feasible due to the computational efficiency of simulating diffraction, leading to the common use of diffraction for determining structure.

However, diffraction only measures the magnitude of the Fourier transform of the electron density, and the lack of phase information limits its use in structural determination \cite{stout1989x, hauptman1991phase}. Assuming positive electron density concentrated around atoms of known atomic numbers often alleviates this ``phase problem" for simple structures \cite{stout1989x,hauptman1991phase}. However, for more complicated systems, such as nanocrystalline systems \cite{billinge2007problem}, systems with local distortions \cite{kalyani2005various,sicolo2020and}, or amorphous systems with broadened, overlapping peaks in the spectra, determining the structure from diffraction can yield non-unique structures. This non-uniqueness indicates structural determination is an ill-posed inverse problem. Simulating spectroscopic data from an atomic structure follows the causality and physics, and thus is a forward problem. The forward problem yields a unique spectrum for a given structure, part of the definition of well-posedness \cite{hasanov2021introduction} associated with forward problems. The inverse problem, however, has no guarantee of a unique solution, and multiple distinct structures could explain the same spectrum, leading to the phase problem. Experimental \cite{van2004coherent} and analytical \cite{stout1989x,hauptman1991phase} approaches exist to address the phase problem, but another common approach is to combine diffraction with additional spectroscopies \cite{clausen1998combined,belyakova2004atomic} that provide similarly incomplete but complementary information about the atomic structure.

Many other spectroscopies used in materials characterization, such as X-ray absorption near-edge structure (XANES) and extended X-ray absorption fine structure (EXAFS), contain detailed structural information but require expensive simulations which preclude the structure solvers commonly applied to diffraction measurements. These spectroscopies are typically interpreted through summary statistics and qualitative descriptors rather than full structural determination. Common fitting algorithms for EXAFS\cite{martini2020pyfitit, newville2001ifeffit} do not propose an atomic structure, but rather fit summary statistics such as bond lengths and coordination numbers. XANES is often reported through statistics such as the edge energy, white-line intensity, and pre-edge energy \cite{wang2021pitfalls,choi2009covalence,togonon2025insights}. It is appropriate to use summary statistics as a single spectroscopy does not contain complete information about the atomic structure. However, summary statistics are also employed because the computational cost of simulating XANES and EXAFS makes iteratively evaluating the forward problem (these simulations) to solve the inverse problem prohibitively expensive. Expanding structural determination solvers to a broader range of spectroscopies requires algorithms that more efficiently leverage structure-to-spectra simulations. 

Most solvers for inverse problems with an expensive forward problem replace the forward problem with a surrogate model. For structural determination, the surrogate model must predict the spectrum corresponding to a given structure, approximating the more accurate simulations of the forward problem at a lower computational cost. Machine learning models satisfy the requirements of quickly and accurately reproducing the complex structure-to-spectra relationship. Specifically, graph neural networks leverage the chemical structure to better approximate properties such as spectra \cite{carbone2019classification,fung2021benchmarking,kwon2024spectroscopy,kharel2025omnixas}. The high computational cost of generating training data from simulations for these models also necessitates deliberate and efficient data sampling. One strategy involves training the model on existing general-purpose datasets such as the Materials Project's XAS spectra\cite{mathew2018high,rankine2020deep}. Published comparisons to experimental data \cite{mathew2018high} and nonphysical artifacts, such as negative X-ray absorption, \cite{carbone2019classification} demonstrate a lack of quantitative accuracy in this dataset, which limits models trained on it. A recent XANES dataset and associated forward model \cite{kharel2025omnixas} may alleviate these issues, but this model has not yet been integrated into a structural determination framework. Additionally, the search space for structural determination is the set of possible structures for a given stoichiometry. For even the largest general-purpose datasets, only a few polymorphs have simulated spectra for most stoichiometries. Another strategy involves training a bespoke model for the materials application at hand. A recent study\cite{kwon2024spectroscopy} uses a trained forward model as the conditional part of a generative model to find a structure that reproduce a given amorphous carbon XANES spectrum. In this work, we introduce ActiveStructOpt, which instead uses active learning to efficiently sample structure-spectra pairs on the fly to train a surrogate model with as few simulations as possible. 

Even with an accurate and efficient model, inverse problem solvers can falter by being overly reliant on an informative starting guess. Local structural determination from diffraction by gradient-based Rietveld refinement requires a starting structure which is close to the target \cite{mittemeijer2013modern}. Without such a starting structure, or a close enough structure for probabilistic methods, ab initio structural determination determines the structure over the global space \cite{racioppi2025powder}. For polycrystalline and nanocrystalline materials, broadening and overlap in the peaks causes ab initio structural determination to remain an active area of research. One approach involves training generative models on large XRD datasets\cite{guo2024diffusion, riesel2024crystal}. Another combines energy minimization and spectroscopic goodness-of-fit objectives in the optimization\cite{racioppi2025powder, meredig2013hybrid}. These global methods have primarily been developed for diffraction. The framework introduced here is designed to maintain performance with uninformative initial structures, both by optionally including energy minimization in the optimization, and by searching over the surrogate model with multiple starts. When available, a Rietveld-refined structure can serve as effective initial structure when fitting other spectroscopies. We benchmark this framework in both cases where the starting structure well approximates the target structure, and in cases where the structures deviate substantially, to test the effectiveness of this framework as a global structure determination method. 

Recognizing the limitations of a single spectroscopy, recent literature has focused on solving the task of determining the structure from multiple spectra. A solver for simultaneously refining a subset of atomic structure parameters has been developed for XRD and EXAFS \cite{binsted1996combined}. Reverse Monte Carlo with simultaneous fitting of EXAFS and diffraction data has also been developed \cite{wicks1995rmc, winterer2025coupling}, though the large number of simulations required for RMC limit the accuracy of EXAFS simulations employed or require high computational expense, as with EXAFS-only RMC approaches \cite{timoshenko2012reverse}. Using multiple X-ray absorption edges from different elements in the system, a model trained from pre-labeled data performed several characterization tasks more accurately than models using single edges \cite{jia2025revealing}. Energy minimization is another objective that can be combined with spectroscopic objectives, and has been combined for diffraction experiments\cite{racioppi2025powder, meredig2013hybrid}, including multiple diffraction spectra \cite{cuillier2024integrating}. A generalized approach to simultaneously fit multiple objectives, including simulations with expensive simulations, would fully take advantage of the complementary information obtained by thermodynamics and multiple spectroscopies. 

ActiveStructOpt is an active learning framework with a chemically informed surrogate model to propose atomic structures using a limited number of simulations, particularly for computationally expensive spectroscopies such as XANES and EXAFS. Using a graph neural network as the surrogate model effectively captures the features of atomic structures to accurately predict the spectra of any given structure. By using active learning, training these GNNs does not need to rely on large pre-existing spectra datasets or prior structural knowledge to generate problem-specific data, greatly expanding its applicability across material domains. We benchmark ActiveStructOpt against a diverse set of material systems representative of practical structure determination problems, including amorphous and crystalline structures, small and large unit cells, and local and global structural determination. We also selected systems with local or long-range disorder, where Rietveld refinement does not contain enough information to fully determine the structure. This method expands the range of spectroscopies accessible for structural determination, allowing for the determination of structures not determinable by diffraction alone.  

\section{Methods}

\subsection{ActiveStructOpt}

\begin{figure}
     \centering
         \includegraphics[width=\textwidth]{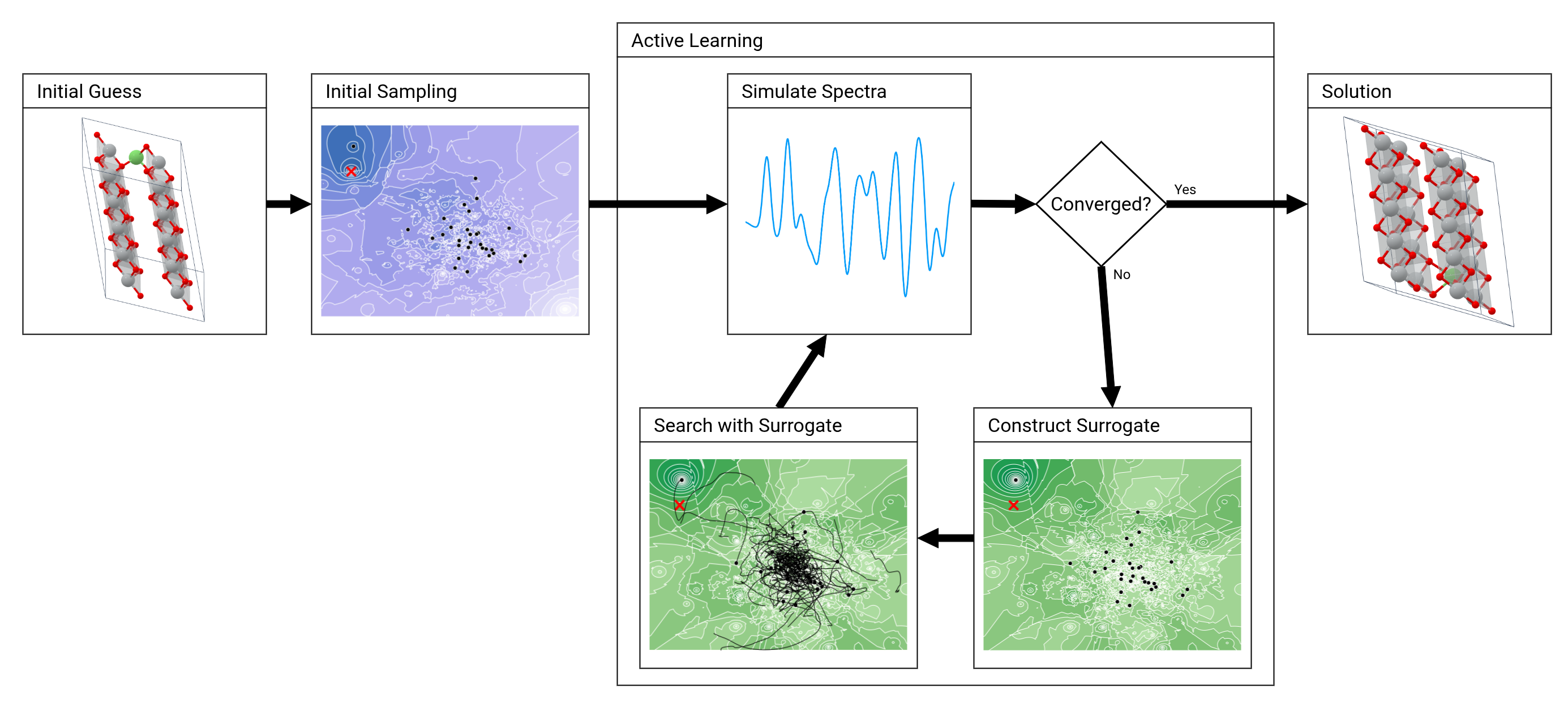}
        \caption{Schematic of the Bayesian optimization loop in ActiveStructOpt, using the structure determination of lithium nickel oxide from an X-ray pair distribution function (PDF) spectrum as an example, following the general surrogate modelling process  \cite{forrester2008engineering}. The blue contour plot maps the goodness-of-fit over a two-dimensional projection of the structural space, where a darker color represents a better goodness-of-fit to the target X-ray PDF (the target structure itself is marked with a red X), and the green contour plots show the same projection with goodness-of-fits predicted by the surrogate model. Black lines represent the multiple optimization traces run over the surrogate model to determine the next candidate structure.}
        \label{fig:overall_method}
\end{figure}

Materials scientists determine structure by finding an atomic structure candidate with a simulated spectrum that best fits the experimental spectrum. Calculating the goodness-of-fit for a candidate involves simulating the spectrum which, for many spectroscopies, can contribute significantly to the computational cost, so an effective method more efficiently uses a small number of simulations. Bayesian optimization facilitates an efficient global search when evaluating the value of the function is expensive \cite{frazier2018bayesian}. Instead of determining the next structure candidate directly from the losses calculated from previous candidates, Bayesian optimization determines the next candidate by optimizing over a constructed surrogate model with a cheaper prediction of this value. The acquisition function used in the surrogate model optimization often performs better when balancing selecting candidates for which calculating the loss will improve the surrogate model accuracy (exploration), and finding a candidate with a better goodness-of-fit (exploitation) \cite{forrester2008engineering}. ActiveStructOpt uses a Bayesian optimization framework as depicted in Figure \ref{fig:overall_method}, the components of which are described below.

\subsubsection{Surrogate Model}

An atomic structure with $N$ atoms in the unit cell can be defined by three components: the atomic numbers of each atom $\mathbf{Z} \in \mathbb{N}^{N}$, the fractional coordinates of each atom $\mathbf{X} \in \mathbb{R}^{3 \times N}$, and the vectors defining the unit cell $\mathbf{L} \in \mathbb{R}^{3 \times 3}$. The sensitivity of this structure representation to spectroscopically irrelevant transformations such as translation or rotation leads to a high-dimensional and multi-modal mapping of the structure to spectra goodness-of-fit. While Gaussian processes are conventionally used as surrogate models for Bayesian optimization, these difficulties lead to Gaussian processes working poorly for this problem, even with a fixed lattice. 

For ActiveStructOpt, we represent the structure as a graph, with atoms as nodes, and the atom-neighbor pairs as edges, up to a cutoff radius of 6-8 \AA\  between neighbors to capture medium-range information. The node encoding includes the atomic numbers, and the edge encoding includes the interatomic distances. This type of transformation has a strong physical basis in spectroscopy; a similar transformation occurs in the simulation of XANES and EXAFS\cite{rehr2010parameter}. 

Graph neural networks learn a latent representation of the atomic structure through message passing functions, which aggregate neighbor information and are updated through regression of the predicted and labeled (simulated) spectra. ActiveStructOpt supports multiple architectures and two GNN architectures are used in this work: TorchMD-NET\cite{tholke2022torchmd}, which is trained from scratch through the MatDeepLearn framework \cite{fung2021benchmarking}, and a pre-trained Orb-v3 model\cite{rhodes2025orb} through MatterTune \cite{kong2025mattertune}. The pre-trained model generally performs better over a variety of benchmarks, but for the amorphous carbon benchmark the large number of atoms requires the use of the faster TorchMD-NET model. The effect of pre-training on the performance of the Orb-v3 model as a surrogate model is explored in Figure S5 in the Supplemental Information. The surrogate model is used to suggest candidate structures that supplement the dataset, and then the surrogate model is retrained with the new data in a process called active learning. Using a GNN as the surrogate model enables accurate spectral prediction from a small dataset, which is necessary for the Bayesian optimization loop. 


\subsubsection{Uncertainty Quantification}

For the surrogate model, we combine multiple GNNs in a k-fold ensemble, with $k = 5$ in all tests in this paper, using the mean as the prediction and the standard deviation as the uncertainty. A test set of $20$ structures are separated to calibrate the uncertainty with a scalar multiplication, as neural network ensembles tend to underestimate uncertainty \cite{tran2020methods}. Acquisition functions that prioritize candidates that improve the surrogate model (exploration) use the uncertainty estimated by this technique. The algorithm for training the surrogate model and calibrating uncertainty is visualized in Figure S1 in the Supplemental Information.

\subsubsection{Constraints}

The search space for atomic structures can be constrained from all numerically possible structures to physically reasonable structures. We exclude structures if any interatomic distance within the structure is less than 85\% of the minimum pairwise distance reported in the Materials Project database\cite{jain2013commentary,horton2025accelerated} for the pair of atomic numbers. Soft, differentiable constraints are implemented to facilitate gradient-based optimization where interatomic distances below the cutoffs penalize the acquisition function using the repulsive term of a Lennard-Jones potential \cite{wang2020lennard}. 

\subsubsection{Sampling}

We benchmark using two sampling strategies: a random perturbation sampler (for fixed unit cells) and a space-group-based sampler (for variable unit cells). The random perturbation sampler perturbs atomic positions uniformly within a specified distance. The sampler rejects and resamples candidate structures that do not satisfy the constraints. The space-group-based sampler\cite{fredericks2021pyxtal} only uses the composition from the initial structure. A random space group is sampled from the space groups possible given this composition and the constraints, weighted by prevalence in the Materials Project\cite{jain2013commentary,horton2025accelerated}. Then the PyXtal\cite{fredericks2021pyxtal} method samples a structure with this space group, subject to the constraints. For all tests, the initial structure and $9$ other sampled structures are used to initially train the surrogate model, and $20$ additional structures are generated for uncertainty calibration. The sampling algorithms are visualized in Figure S2 in the Supplemental Information.

\subsubsection{Acquisition Function}

The acquisition function we use is a lower confidence bound \cite{forrester2008engineering} modified to handle spectroscopic data, which combines an exploitation objective (mean absolute error between the predicted and target spectra) and an exploration objective (scaled average uncertainty of the predicted spectra). The uncertainty term is omitted in the final iterations to prioritize exploitation when the model is sufficiently accurate. Soft interatomic distance constraints are incorporated into the acquisition function, increasing efficiency by reducing the search space and screening unphysical candidate structures that may cause simulations to fail. 

\subsubsection{Optimization}

For the benchmarks in this paper, we optimize over the surrogate model using the AdamW algorithm \cite{loshchilov2017decoupled}, though different optimizers can be used within this framework. Gradients of the acquisition function with respect to structural parameters, $\mathbf{L}$ when applicable and $\mathbf{X}$, propagate through the graph neural network construction, allowing for gradient-based optimizers. Several independent optimization trajectories start from structures selected from the dataset and the initial sampling distribution. The structure along the trajectories that has the best acquisition function becomes the next candidate structure. The simulator then computes the associated spectrum and the structure-spectra pair is added to the dataset. The surrogate model is retrained to include the new data, and this active learning loop repeats until reaching a predefined maximum number of simulations. The algorithm for the optimization is depicted in Figure S3 in the Supplemental Information.

\subsubsection{Benchmark Configurations}

We benchmark ActiveStructOpt over a variety of material cases with a budget of 200 simulation calls, including 10 randomly sampled structures for the initial training dataset, and 20 randomly sampled structures for uncertainty quantification. We perform 130 iterations in a balanced exploration-exploitation mode where the uncertainty quantification biases the optimization towards structures with high prediction uncertainty. We perform the last 40 iterations in a pure exploitation mode, solely optimizing the goodness-of-fit between the predicted and target spectra. The Supplemental Information contains complete configuration settings for the different benchmarks. 

\subsection{Baseline Methods}

\subsubsection{Reverse Monte Carlo}

We tested Reverse Monte Carlo (RMC), a commonly used algorithm for structural determination, over the benchmarks for comparison. For each step, a single atom is randomly perturbed with a displacement up to $0.1$ \AA, and a Metropolis-Hastings criterion determines the acceptance or rejection of this candidate structure, given a temperature parameter of $0.0025$ and the goodness-of-fit of the simulated spectrum \cite{mcgreevy2001reverse}. Interatomic distance constraints also reject unphysical structures in addition to the Metropolis-Hastings condition, but these rejections do not cost a simulation. We ran each RMC benchmark with many more simulations than ActiveStructOpt in order to check the difference in simulation efficiency. 

\subsubsection{DiffPy}

We also compare ActiveStructOpt against the PdfFit2 algorithm \cite{farrow2007pdffit2} from the DiffPy-CMI suite for the X-ray PDF benchmarks. PdfFit2 uses the derivative of the goodness-of-fit with respect to structural parameters in a Levenberg-Marquardt optimization. PdfFit2 optimizes fractional atomic coordinates, and for variable lattice benchmarks, lattice angles and lengths. Due to occasional numerical instability with Levenberg-Marquardt \cite{de2020stability}, at each step, lattice angles are reduced modulo 180\degree\  and the fractional coordinates are reduced modulo 1. When this method fails, typically due to singular matrices, we slightly perturb the initial structure and restart the optimization from the perturbed structure. The X-ray PDF simulations used in PdfFit2 are slightly different from those used for the other methods, so for the DiffPy benchmarks, the target spectra are simulated with PdfFit2-accessible settings. 

The DiffPy method relies on analytic derivatives which are not available for many of the spectroscopies targeted by ActiveStructOpt, such as XANES and EXAFS. DiffPy thus represents an idealized benchmark and motivates the use of approximate gradient-based optimization with a surrogate model.

\subsubsection{Bayesian Optimization with Gaussian Processes}

We used the BOTorch framework \cite{balandat2020botorch} to evaluate a standard Bayesian optimization algorithm using Gaussian processes to compare on the same benchmark as a baseline. For each iteration, this method trains a Gaussian process and uses it to select a candidate structure by optimizing the expected improvement acquisition function. We omitted nonlinear interatomic distance constraints due to the complexity of implementation. Using fractional coordinates as the input space allows for consistent bounds independent of the lattice. Due to poor performance and high computational cost on even the fixed-lattice high temperature perturbations benchmark, we did not test this method for each of the remaining benchmarks.

\subsection{Benchmark Metrics}

\subsubsection{Goodness-of-Fit}

We quantify the goodness-of-fit by the mean squared error (MSE), given the approximately Gaussian errors in many spectroscopies. To evaluate efficiency, we plot MSE as a function of the number of simulations, as ActiveStructOpt aims to reduce the number of simulations required for structural determination. Since users could stop ActiveStructOpt upon achieving a particular goodness-of-fit, the performance after a particular number of simulations is defined as the lowest MSE achieved to that point. Accordingly, the performance curves show the accumulated minimum MSE found within a particular number of simulations. For multi-structure benchmarks, the median of the cumulative minimum MSEs is reported. As such, the performance curve reflects the performance achieved by at least half of the structures measured, using a particular number of simulations. To compare efficiency, we report the number of simulations required by other methods (Bayesian optimization with Gaussian processes, reverse Monte Carlo, and DiffPy) to achieve the same median performance. 

\subsubsection{Structure Matching}

We attempted to match the target and optimized structures using the pymatgen \cite{ong2013python} StructureMatcher, which converts structures to their unique reduced Niggli cells, and then aligns pairs of structures with translation, rotation, and inversion operations. Although inversion is not a symmetry element of enantiomorphic space groups, chirality does not affect the standard X-ray PDF \cite{tanaka2008right}. By the default settings in pymatgen, a pair of structures ``match" if, after alignment, the root mean squared displacement between atoms is less than $0.3$ times the average free length per atom. For lattice variable benchmarks, lattice fractional lengths must agree within $0.2$ and the lattice angles must agree within $5$ degrees for the structures to ``match". A non-matching optimized structure may have a better goodness-of-fit than a structure that (imperfectly) matches, so the reported match percentage likely underestimates the algorithm's performance. 

\section{Results}

\subsection{X-ray Pair Distribution Function}

\begin{figure}
     \centering
     \begin{subfigure}[b]{1.0\textwidth}
         \centering
         \includegraphics[width=\textwidth]{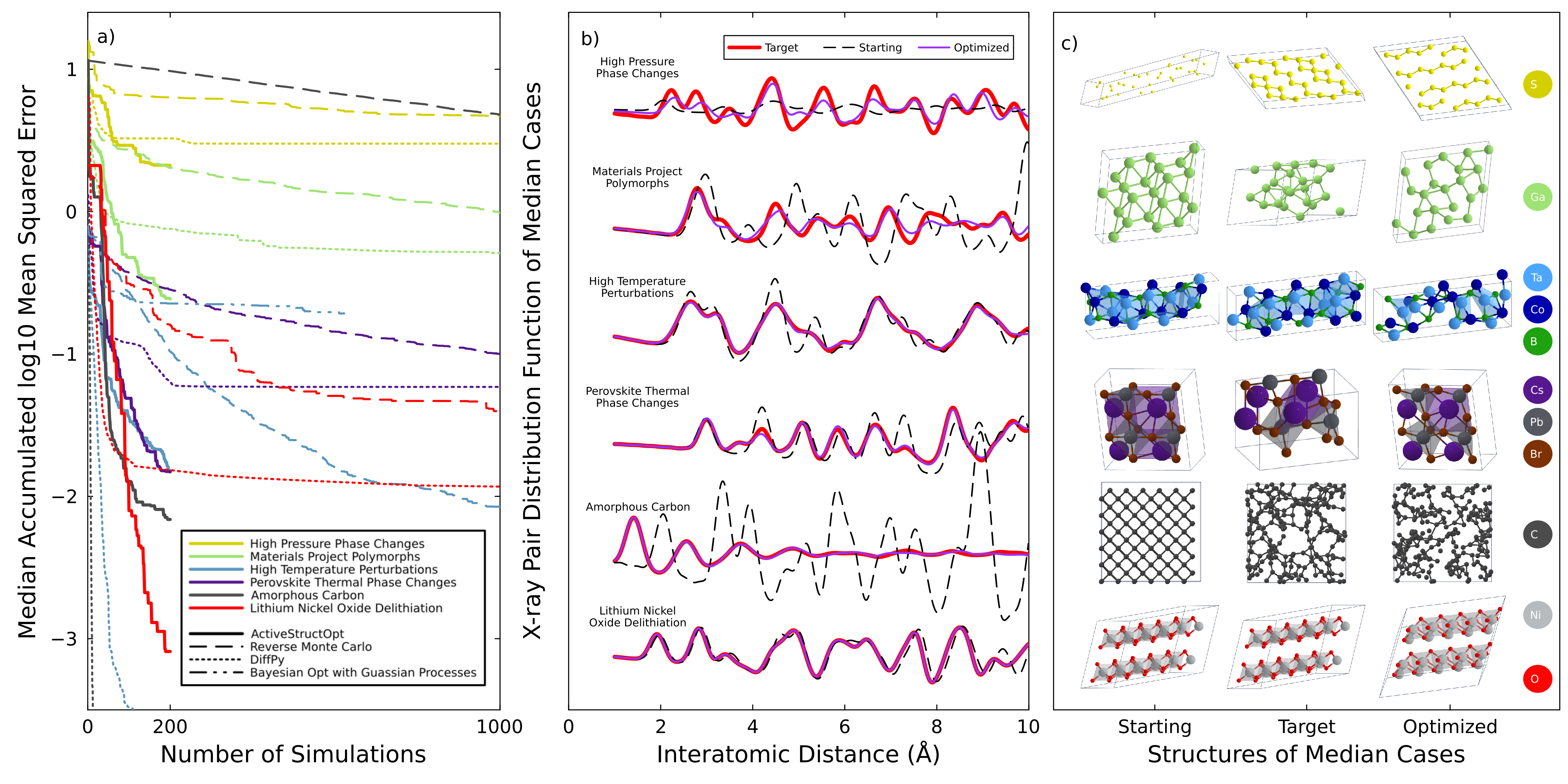}
         \label{fig:xpdf_combined_sf}
     \end{subfigure}
        \caption{Results for the X-ray pair distribution function benchmarks. a) Performance as measured by the the median over the test cases of the best mean squared error seen after a number of simulations. ActiveStructOpt (solid), reverse Monte Carlo (dashed), DiffPy (dotted), and Bayesian Optimization with Gaussian Processes (dashdot) methods are compared. b) Simulated X-ray PDFs of the starting (black, dashed), target (red, solid), and optimized (purple, solid) structures for the median performing case in each benchmark. Structural determination minimizes the mean squared error between the target and optimized PDFs. c) Crystal Toolkit \cite{horton2023crystal} visualizations of the starting (left), target (middle), and optimized (right) structures. Legends indicate the color associated with each atomic species.}
        \label{fig:xpdf_combined}
\end{figure}

We first evaluate ActiveStructOpt on structure determination from X-ray pair distribution function spectra. PDF serves as a useful benchmark to perform over a wide range of materials systems, since PDF simulations are cheap and several existing methods exist for structure determination for comparison. The X-ray pair distribution function was simulated with DiffPy \cite{juhas2017diffpy}, between 1 and 10 \r{A}. Table \ref{tab:pdf} and Figure \ref{fig:xpdf_combined} show a summary of the results of the PDF benchmarks.


\begin{center}
\begin{table}
\resizebox{\textwidth}{!}{
    \begin{tabular}{|| p{5.75cm} || p{1.4cm} | p{1.0cm} | p{1.6cm}| p{2.0cm} || p{2.25cm} | p{2.5cm} |  p{2.5cm} | p{1.9cm} ||} 
     \hline
     \small Task & \small Variable Lattice & \small \# of Atoms & \small \# of Problems & \small Surrogate Model & \scriptsize Log10 Median MSE After 200 Iterations ($\downarrow$) & \scriptsize RMC \mbox{Iterations} For Same MSE ($\uparrow$) & \scriptsize DiffPy \mbox{Iterations} For Same MSE ($\uparrow$) & \scriptsize Percent Matching Structures ($\uparrow$) \\ [0.5ex] 
     \hline\hline
      \small High Pressure Phase Changes & Yes & 12-32 & 13 & Orb-v3 & 0.32 & \textgreater 10,000 & \textgreater 10,000 & 23\% \\
     \hline
     \small Materials Project Polymorphs & Yes & 20-30 & 100 & Orb-v3 & -0.61 & \textgreater 10,000 & \textgreater 10,000 & 3\%\\
     \hline
     \small High Temperature Perturbations & No & 20-30 & 100 & Orb-v3 & -1.82 & 616 & 29 & 34\%\\ 
     \hline
     \scriptsize Perovskite Thermal Phase Changes & Yes & 20 & 20 & Orb-v3 & -1.83 & \textgreater 10,000 & \textgreater 10,000 & 50\% \\ 
     \hline
     \small Amorphous Carbon & No & 216 & 30 & \scriptsize TorchMD-NET & -2.16 & \textgreater 9,000 & 8 & 0\% \\
     \hline
     \small Lithium Nickel Oxide Delithiation & Yes & 60-72 & 13 & Orb-v3 & -3.09 & \textgreater 10,000 & \textgreater 10,000 & 100\% \\ 
     \hline
    \end{tabular}
}
\caption{Problem setup and results of the X-ray PDF benchmarks. }
\label{tab:pdf}
\end{table}
\end{center}

We constructed several benchmarks representing unknown phase transitions with a known stoichiometry and an initial structure that shares little structural similarity to the target in either the atomic positions or lattice parameters. The High Pressure Phase Changes benchmark is 13 experimentally verified monatomic structures at 300 GPa from the HEX database \cite{giannessi2024database}, with the starting structures being the corresponding ambient pressure structures. For the Materials Project Polymorphs benchmark, initial and target structures are the 100 pairs of structures from the Materials Project \cite{jain2013commentary,horton2025accelerated} that have identical stoichiometry but maximally different radial distribution functions. For the Perovskite Thermal Phase Changes benchmark, target and initial structures are selected based on experimentally observed phase transitions for perovskites of the formula \ce{CsMX3} where M = \ce{Pb}, \ce{Sn} and X = \ce{Cl}, \ce{Br}, \ce{I}\cite{alaei2021polymorphism}, with 20 pairs of structures of the corresponding crystal systems pulled from the Materials Project\cite{jain2013commentary,horton2025accelerated}.

ActiveStructOpt can also be applied to disordered structures, so we constructed two benchmarks to test this application. For these benchmarks, the lattice parameters are held constant. For the High Temperature Perturbations benchmark, we randomly selected 100 random structures from the Materials Project\cite{jain2013commentary,horton2025accelerated} (between 20-30 atoms per unit cell) as starting structures and perturbed these structures with molecular dynamics in a constant-temperature, constant-volume (NVT) ensemble \cite{bitzek2006structural} \cite{larsen2017atomic} using the MACE MP medium model \cite{batatia2023foundation} to generate the target structures. For the Amorphous Carbon benchmark, we target 30 structures in a dataset from a publication studying structural determination from XANES \cite{kwon2024spectroscopy}, starting from diamond structures expanded to the same volume. 

The Lithium Nickel Oxide Delithiation benchmark represents a system with more subtle phase changes. The target \ce{Li_xNiO2} structures where $x \le 0.6$ are from the reported XRD structures \cite{hirano1995relationship}, not accounting for non-stoichiometric Ni substitutions in the Li sites, and reducing the unit cell by rounding $x$ to the nearest 1/20. We construct the starting structures by removing the appropriate number of lithium atoms from the fully lithiated ($x = 1$) XRD-determined structure, without relaxation.

In all benchmarks, ActiveStructOpt outperforms reverse Monte Carlo with the same budget. In the disordered structure benchmarks, derivative-enabled methods such as DiffPy may be more efficient, but for many spectroscopies such derivatives are not available. With the exception of some of the global optimization benchmarks (High Pressure Phase Changes and Materials Project Polymorphs), the median case yields deviations less than or equal to the typical errors associated with experimental measurements. The Supplemental Information contains more details about each benchmark and the associated results. 

\subsection{X-ray Absorption Spectroscopy}

Simulation settings are available in the Supplemental Information for all XAS simulations described below. All XAS simulations are through FEFF, a Green's function based simulator \cite{rehr2010parameter}.

\subsubsection{Amorphous Carbon}

Powder X-ray diffraction contains limited information for disordered materials, and the pair distribution function incompletely describes the structure (see the angular distribution discrepancy in well-matched PDFs in Figure S8 in the Supplemental Information). Shelby's Fundamental Law of Structural Models states that “No model can be considered to be valid unless that model can explain ALL of the available data,” \cite{shelby2020introduction}, highlighting the importance of quantitative structural determination from other spectroscopies, such as X-ray absorption spectroscopy. For this amorphous carbon benchmark, we computed the EXAFS with a cutoff radius of 8 \AA\  over a k-range of 3-12 \AA$^{-1}$. As in the earlier X-ray PDF benchmark for amorphous carbon, the target structure is amorphous phase \cite{kwon2024spectroscopy}, and the starting structure is the diamond phase. While the optimized structure roughly matches the target EXAFS curve especially considering the extreme deviation of the starting structure, the agreement is not within typical experimental error. Reverse Monte Carlo does not approach the spectra matching performance of even the random generated structures when given twice the simulation budget for this problem.

\begin{figure}
     \centering
     \begin{subfigure}[b]{1.0\textwidth}
         \centering
         \includegraphics[width=\textwidth]{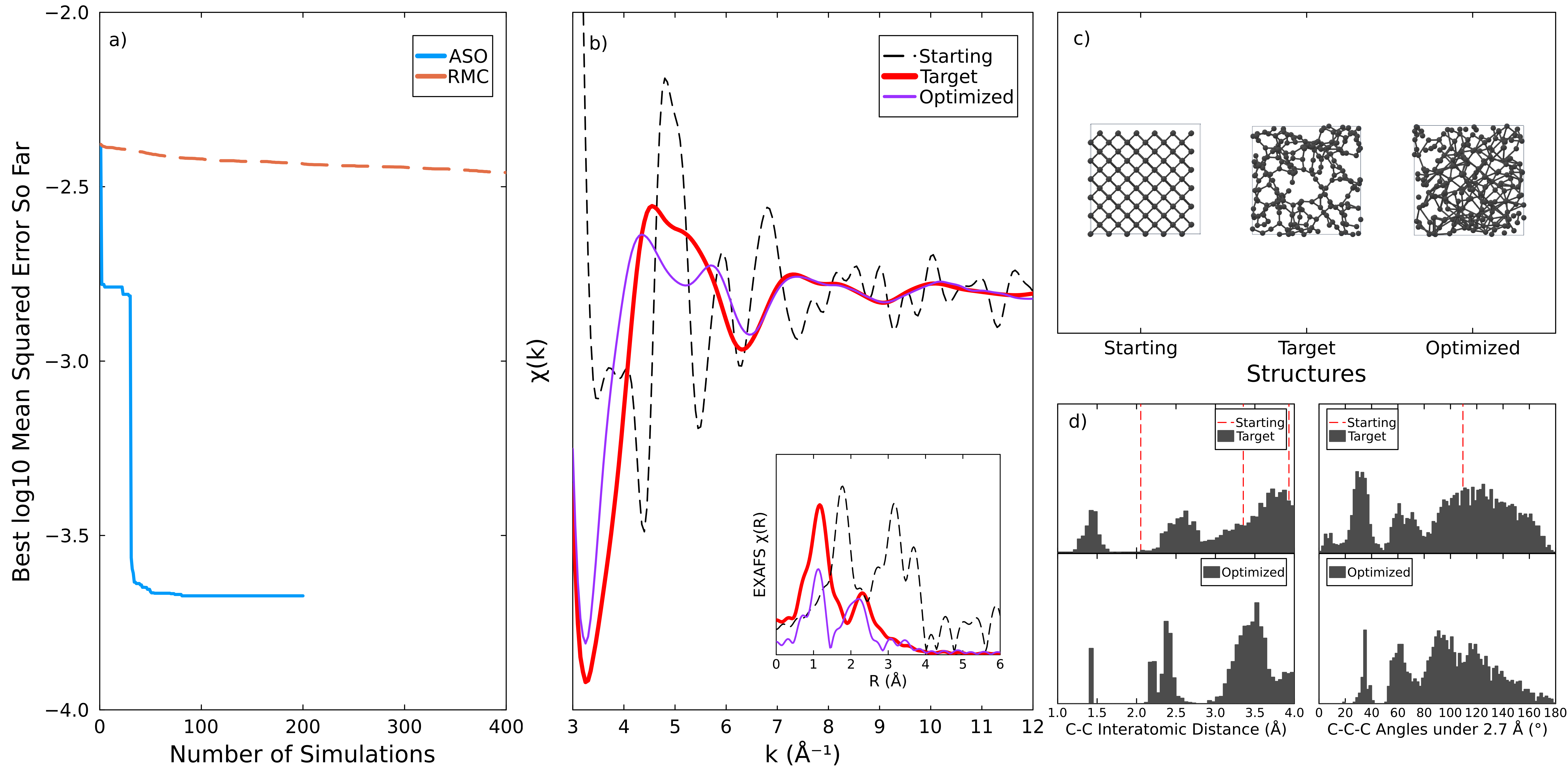}
         \label{fig:amorph_feff_sf}
     \end{subfigure}
        \caption{Results for the amorphous carbon EXAFS test. a) Performance of ActiveStructOpt (blue, solid) and reverse Monte Carlo (orange, dashed) as measured by the the median over the test cases of the best mean squared error seen for a given number of simulations. b) Simulated EXAFS of the starting (black, dashed), target (red, solid), and optimized (purple, solid) structures. Inset shows the same information in R-space. c) Crystal Toolkit \cite{horton2023crystal} visualizations of the starting (left), target (middle), and optimized (right) structures. d) Histograms comparing the distributions of interatomic neighbor distances (left) and angular distributions between neighbors less than 2.7 angstroms separated (right). Dashed lines indicate values for the starting diamond structure. }
        \label{fig:amorph_feff}
\end{figure}

\subsubsection{Lithium Nickel Oxide Jahn-Teller Distortions}

\begin{figure}
     \centering
     \begin{subfigure}[b]{1.0\textwidth}
         \centering
         \includegraphics[width=\textwidth]{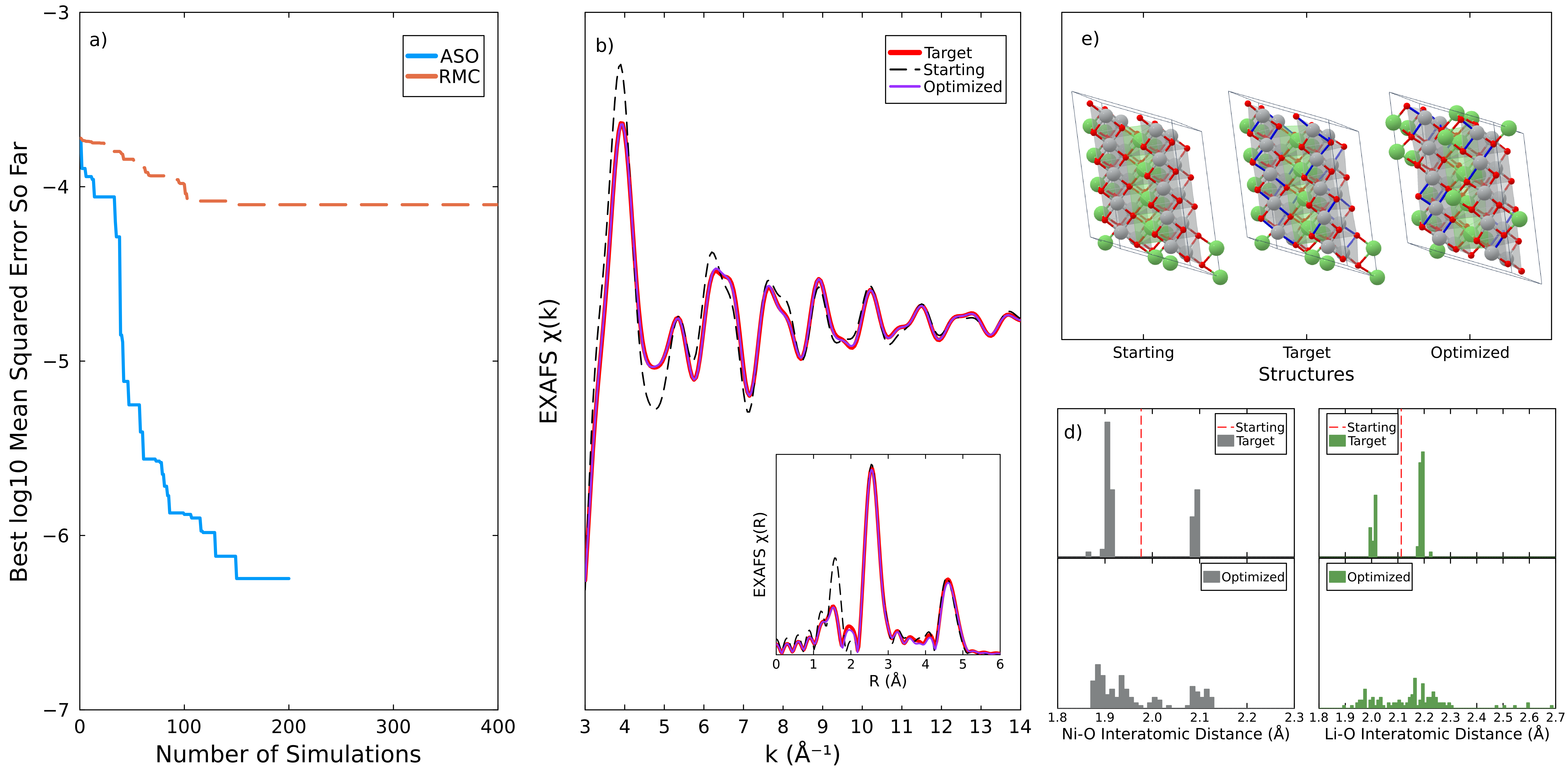}
         \label{fig:lno_exafs_mult_sf}
     \end{subfigure}
        \caption{Results for the lithium nickel oxide Jahn-Teller distortions EXAFS test. a) Performance of ActiveStructOpt (blue, solid) and reverse Monte Carlo (orange, dashed) as measured by the the median over the test cases of the best mean squared error seen in a number of simulations. b) Simulated nickel EXAFS of the starting (black, dashed), target (red, solid), and optimized (purple, solid) structures. Inset indicates the real-space visualization of the EXAFS. c) Crystal Toolkit \cite{horton2023crystal} visualizations of the starting, target, and optimized structures. Dark blue bonds indicate Ni-O distances longer than 2 \AA. d) Histograms of the nickel--oxygen and lithium--oxygen nearest neighbor distances. Dashed lines indicate values for the starting structure.}
        \label{fig:lno_exafs_mult}
\end{figure}

\begin{figure}
     \centering
     \begin{subfigure}[b]{1.0\textwidth}
         \centering
         \includegraphics[width=\textwidth]{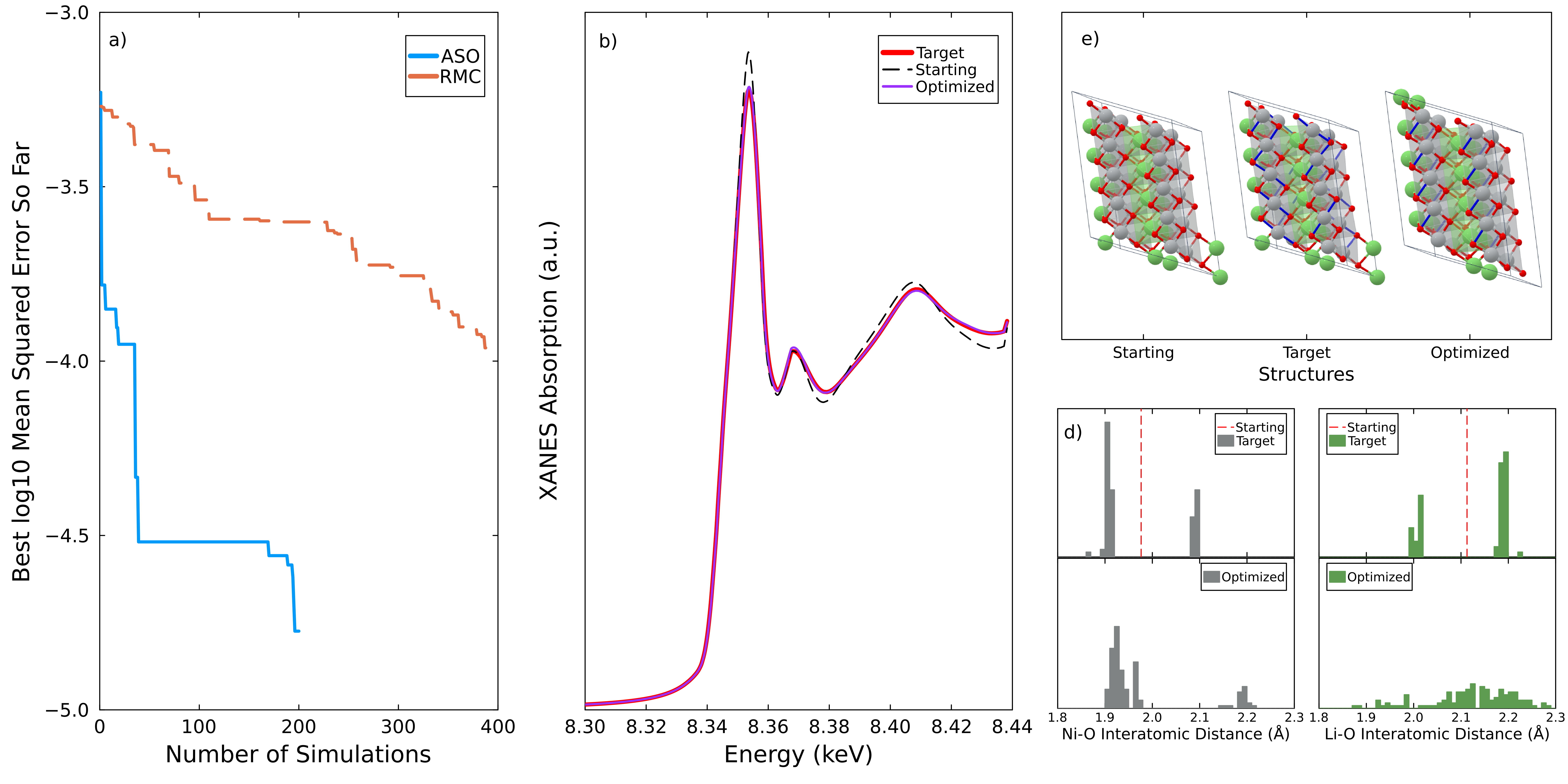}
         \label{fig:lno_xanes_mult_sf}
     \end{subfigure}
        \caption{Results for the lithium nickel oxide Jahn-Teller distortions XANES test. a) Performance of ActiveStructOpt (blue, solid) and reverse Monte Carlo (orange, dashed) as measured by the the median over the test cases of the best mean squared error seen in a number of simulations. b) Simulated nickel XANES of the starting (black, dashed), target (red, solid), and optimized (purple, solid) structures. c) Crystal Toolkit \cite{horton2023crystal} visualizations of the starting, target, and optimized structures. Dark blue bonds indicate Ni-O distances longer than 2 \AA. d) Histograms of the nickel--oxygen and lithium--oxygen nearest neighbor distances. Dashed lines indicate values for the starting structure. }
        \label{fig:lno_xanes_mult}
\end{figure}

When originally synthesized and also in more modern measurements, powder XRD indicated that \ce{LiNiO2} has a $R\mkern1mu\overline{\mkern-1mu3\mkern-1mu}\mkern1mu m$ space group, and equidistant Ni-O bonds in the octahedra \cite{dyer1954alkali,hirano1995relationship,kalyani2005various}. \ce{NaNiO2}, by contrast, has a C2/m monoclinic structure, with distorted Ni-O octahedra \cite{phillips2025collinear}. Under standard assumptions for oxygen and alkali metals, the oxidation state of nickel in both structures is 3+, and thus the Jahn-Teller theorem predicts the octahedral nickel to be distorted irrespective of spin state \cite{miessler2004inorganic}. The lack of Jahn-Teller distortion in \ce{LiNiO2} was originally ascribed to the Ni$^{2+}$ intermixing in the lithium layer \cite{dutta1992chemical}. However, an EXAFS study found two distinct Ni-O distances in the Fourier-transformed distribution \cite{rougier1995non}, which was originally reported as two distinct peaks, but more recent work by the same author reports the distortion causing a decrease in intensity in the Ni-O peak \cite{mansour2000situ}. The decrease in intensity in the Ni-O peak is likely more robust to EXAFS processing parameters, and is not explained by thermal effects on the $R\mkern1mu\overline{\mkern-1mu3\mkern-1mu}\mkern1mu m$ structure \cite{lomeli2025negative}. Configurations of Jahn-Teller distorted Ni-O octahedra have been found to be energetically favorable in DFT simulations \cite{chen2011first,sicolo2020and}. One explanation for such behavior is that the Jahn-Teller distortions are oriented within the system with no long-range order, perhaps thermally transitioning \cite{sicolo2020and,genreith2024probing}. To model one possible structure, our target spectra is the EXAFS simulated from a structure with randomly aligned Jahn-Teller distortions. The starting structure is the $R\mkern1mu\overline{\mkern-1mu3\mkern-1mu}\mkern1mu m$ structure determined by XRD. This models a situation where the scientist measures both the EXAFS and XRD and uses the EXAFS to further refine the XRD-determined structure (with fixed lattice parameters) to account for local disorder. The optimized spectrum is much closer than the starting structure. Structurally, the Ni-O bond distance distributions do not perfectly match and the longer Ni-O bonds are not antipodal, further showing that EXAFS is not enough to fully determine the structure. Though the optimization is in k-space, ActiveStructOpt matches the general features of the Fourier-transformed R-space spectra (inset of Figure \ref{fig:lno_exafs_mult}(b)).  We also tested the same case with XANES simulated with FEFF. At this level of theory, the differences between the starting and target spectra is relatively small, but ActiveStructOpt found a structure that matches the target spectra quite closely (Figure \ref{fig:lno_xanes_mult}), and splits the Ni-O bond distances, though in a different manner than in the target structure. 

\pagebreak

\subsection{Structure Determination for Multi-spectroscopy}

Structural determination from a single spectroscopy is an ill-posed inverse problem, as structural solutions are often not unique. This is evident from benchmarks discussed earlier, where pairs of optimized and target structures that are dissimilar according to the pymatgen StructureMatcher but have close X-ray PDFs (for example the median perovskite phase transition case in Figure \ref{fig:xpdf_combined}). In the amorphous case, even the best match to the X-ray PDF has a different angular distribution (Figure S8), indicating the optimized and target structures have significant differences despite both solving the problem. As a theoretical example, enantiomorphic crystal structures that differ only in chirality are indistinguishable by standard X-ray PDF measurements \cite{tanaka2008right}. The limitations above are not specific to X-ray PDFs. A common approach to address these limitations is to measure complementary spectroscopies for the same material. However, quantitatively combining the information from complementary spectroscopies for structural determination remains rare. Incorporating complementary spectroscopies or energetic constraints may serve to constrain the inverse problem and improve the solution uniqueness. We demonstrate multiple objectives with benchmarks in two cases: Jahn-Teller distorted lithium nickel oxide (Ni XANES and EXAFS), and the Materials Project polymorphs benchmark (PDF and energy). An additional example with structures from the perovskite benchmark optimized with PDF and Cs EXAFS is provided in the Supplemental Information.

\subsubsection{Lithium Nickel Oxide Jahn-Teller Distortions}

\begin{figure}
     \centering
     \begin{subfigure}[b]{1.0\textwidth}
         \centering
         \includegraphics[width=\textwidth]{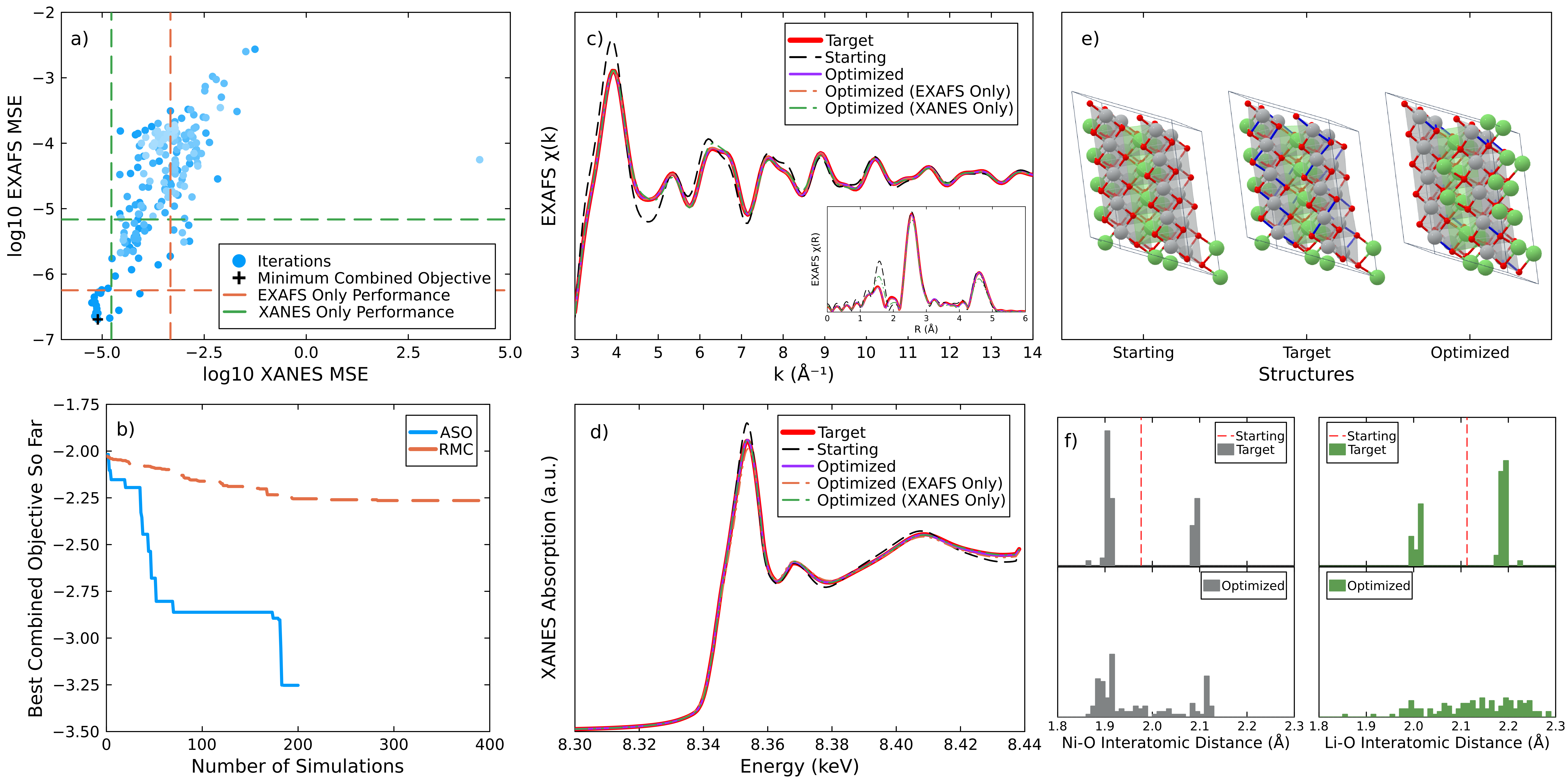}
         \label{fig:lno_mult_sf}
     \end{subfigure}
        \caption{Results for the lithium nickel oxide Jahn-Teller distortions multispectroscopy test. a) The performance as measured by mean squared error (MSE) of each spectroscopy, with darker points indicating later iterations. The performance of the structure that minimizes the combined objective is indicated as a black cross. b) The combined objective of the XANES and EXAFS, as is minimized in the optimization phase, as a function of the number of simulations run for ActiveStructOpt (blue, solid) and reverse Monte Carlo (orange, dashed). c) Simulated nickel EXAFS of the starting (black, dashed), target (red, solid), and optimized (purple, solid) structures. d) Simulated nickel XANES of the starting (black, dashed), target (red, solid), and optimized (purple, solid) structures. e) Crystal Toolkit \cite{horton2023crystal} visualizations of the starting, target, and optimized structures. Dark blue bonds indicate Ni-O distances longer than 2 \AA. f) Histograms of the nickel--oxygen and lithium--oxygen nearest neighbor distances. Dashed lines indicate values for the starting structure. }
        \label{fig:lno_mult}
\end{figure}

ActiveStructOpt was used to simultaneously optimize the nickel XANES and EXAFS, starting from the XRD-derived structure and targeting the spectra of a structure with randomly oriented Jahn-Teller distortions, as in the XANES-only and EXAFS-only benchmarks described previously. The multiple objectives were incorporated into the acquisition function as a weighted sum, scaled such that contribution from each spectra is roughly equal. While a Pareto-aware acquisition function \cite{yong2025bayesian} may improve performance, a simple weighted sum still identified multiple Pareto-optimal structures (Figure \ref{fig:lno_mult}(a)). The structure minimizing the acquisition function is selected as the solution, although any point on the Pareto front may be valid depending on the objective prioritization. Joint optimization outperforms single-spectroscopy cases (XANES or EXAFS only) on their respective objectives with the same budget (Figure \ref{fig:lno_mult}(a)). The optimized EXAFS and XANES spectra are within typical experimental uncertainties (Figure \ref{fig:lno_mult}(c,d)), implying that this combination of spectroscopies, simulated at this level of theory, cannot distinguish between the target structure and the optimized structure under typical measurement conditions. However, the target and optimized structures are different, as illustrated by location of longer Ni-O bonds (Figure \ref{fig:lno_mult}(e)) and the nearest neighbor interatomic distances (Figure \ref{fig:lno_mult}(f)). Although nickel--oxygen distance splitting is present in some octahedra in the optimized structure, there are also nickel--oxygen distances in between the split distances for some octahedra. The lithium--oxygen nearest neighbor distances show greater mismatch, though this is expected, as nickel X-ray absorption spectroscopy is less sensitive to the lithium-oxygen environment. Recent findings \cite{jacquet2024fundamental} corroborate that X-ray absorption spectroscopy alone cannot unambiguously distinguish between some proposed lithium nickel oxide structures, specifically Jahn-Teller distorted and bond disproportionation models. Distinguishing between the target and optimized structures requires additional characterization techniques. The addition of X-ray diffraction alone is unlikely to distinguish between the optimized and target structures, as the differences in peak heights of their simulated XRD powder patterns \cite{ong2013python} are less than 0.05\% of the strongest peak height. Establishing that a set of spectroscopies contains sufficient information for unambiguous structural determination remains an open challenge.

\pagebreak

\subsubsection{Materials Project Polymorphs with Energy}

Generative models effectually constrain the structural search space through the unconditional component which encodes a prior of chemical plausibility or stability. A similar effect can be included in ActiveStructOpt by including energy minimization in the optimization. The Materials Project polymorphs benchmark, described earlier, was chosen to test the inclusion of energy minimization as an objective, as the sample size (100 structures) is large, polymorphs can have substantially different energies, and the random selection of starting and target structures leads to an unbiased sampling of energy changes between starting and target structures. The Orb-v3 pre-trained model \cite{rhodes2025orb} predicts the energy used in the optimization. No active learning was used to update the energy model, though in principle DFT energies could be calculated for the candidate structures and used to update the model. A coarse grid search performed over a subset of the dataset was used to select the weighting of the energy relative to the PDF objective. The optimal weighting is dependent upon metastability or amorphous character of the target structure, as well as temperature and pressure effects. While including an energy constraint did not substantially change the overall matching of the PDF spectra, the similarity between the optimized and target structures did improve. The amount of target and optimized structures that match according to the Pymatgen StructureMatcher improved from 3\% to 12\%.  Energy should be included in structural determination except in specific non-equilibrium cases, as the measured material is expected to be globally stable or metastable, and thus at a local energy minimum. 

\begin{figure}
     \centering
     \begin{subfigure}[b]{1.0\textwidth}
         \centering
         \includegraphics[width=\textwidth]{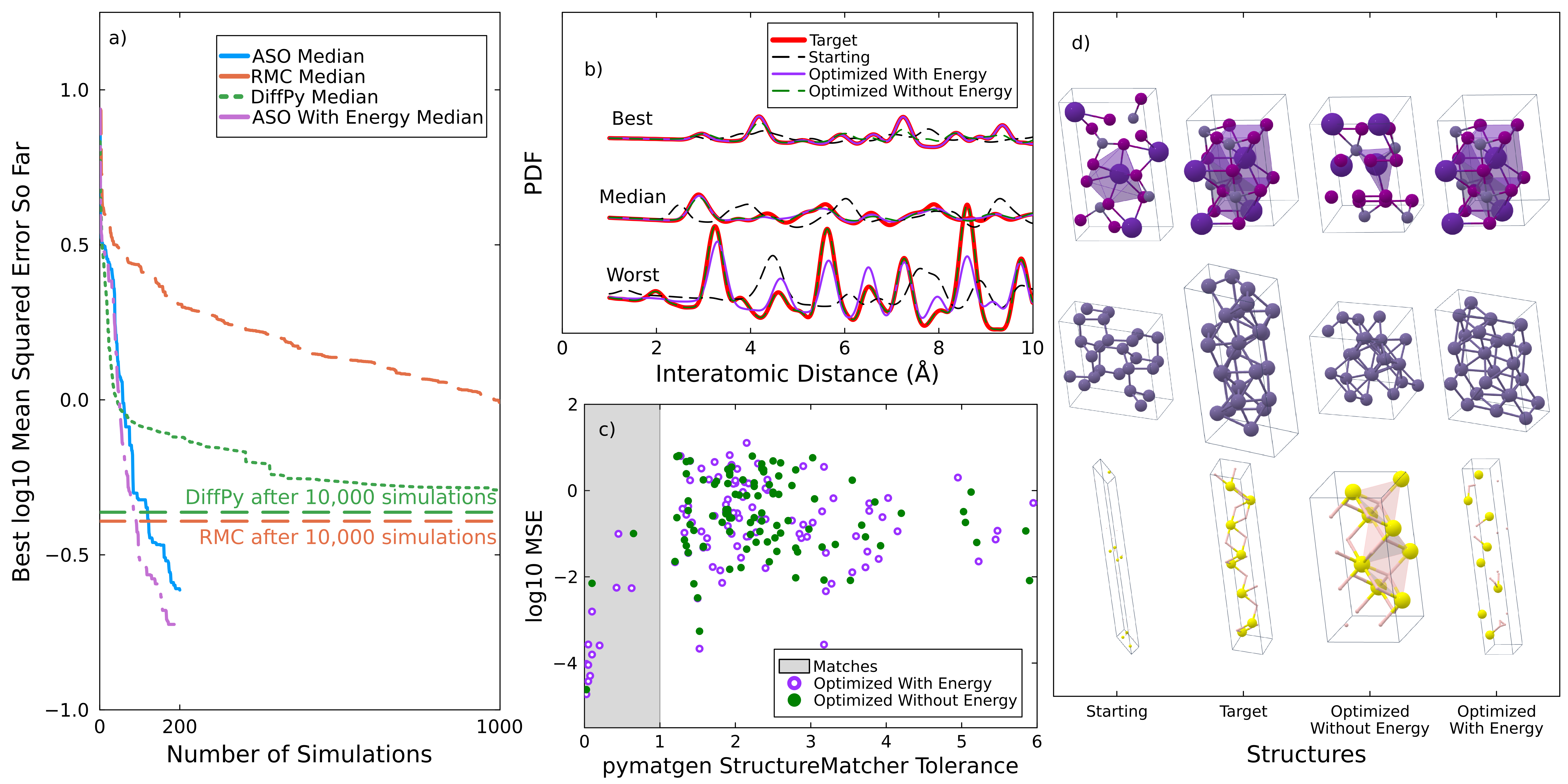}
         \label{fig:mp_energy_sf}
     \end{subfigure}
        \caption{Results for the Materials Project polymorphs with energy test. a) Performance of ActiveStructOpt (blue, solid), reverse Monte Carlo (orange, dashed), DiffPy (green, dotted), and ActiveStructOpt with energy minimization (purple, dashdot) as measured by the the median over the test cases of the best mean squared error seen in a number of simulations. b) Simulated X-ray PDFs of the starting (black, dashed), target (red, solid), optimized with energy (purple, solid), and optimized without energy (green, dashed) structures. Structural determination minimizes the mean squared error between the target and optimized PDFs. c) Scatter plot of the final MSE performance for each structure, compared against the tolerance factor (multiplied by default parameters) required to fit the optimized and target structures with the pymatgen StructureMatcher. Less than 1 on the x-axis is considered a matched structure. Structures determined without energy minimization (green, filled) and with energy minimization (purple, open) are shown. d) Crystal Toolkit \cite{horton2023crystal} visualizations of the starting (left), target (middle left), optimized without energy (middle right), and optimized with energy (right) structures. }
        \label{fig:mp_energy}
\end{figure}



%

\pagebreak

\section{Conclusion}


We have introduced ActiveStructOpt, an active learning framework that treats spectroscopic structure determination as a surrogate-guided inverse problem. By integrating graph neural network surrogates with Bayesian optimization, ActiveStructOpt reduces the number of expensive forward simulations required to accurately determine structure from spectra. Across a diverse suite of benchmarked materials systems and spectroscopic methods, the framework consistently outperforms reverse Monte Carlo and other baseline methods. By actively learning problem-specific surrogate models on the fly, ActiveStructOpt eliminates the reliance on large precomputed spectral databases to perform data-driven structure determination. The framework also naturally accommodates multiple objectives, enabling simultaneous refinement against complementary spectroscopies as well as energetic constraints. We find that including energy as an auxiliary objective improves structural plausibility without degrading spectral agreement, and multi-spectroscopy optimization demonstrably reduces ambiguity relative to single-spectrum fitting.

Importantly, our results reinforce a central challenge in structural science which is that fitting a single spectrum rarely guarantees structural uniqueness. ActiveStructOpt provides a practical computational approach to addressing this ill-posedness by integrating complementary spectroscopic objectives within a unified optimization framework. However, establishing the sufficiency of a given set of spectroscopies for unique structural determination remains an open question and an important direction for future research. As forward simulations and foundation models continue to improve, ActiveStructOpt offers a scalable route toward automated, multi-modal materials characterization and opens new opportunities for resolving complex and disordered structures that lie beyond the reach of diffraction alone.

\section{Data Availability}

The python package ActiveStructOpt can be found at \href{https://github.com/Fung-Lab/ActiveStructOpt}{https://github.com/Fung-Lab/ActiveStructOpt}.

\subsection{Author Contributions}

The manuscript was written through contributions of all authors. All authors have given approval to the final version of the manuscript.  

\subsection{Funding Sources}

This material is based upon work supported by the National Science Foundation Graduate Research Fellowship under Grant No. DGE-2039655. Any opinions, findings, and conclusions or recommendations expressed in this material are those of the authors and do not necessarily reflect the views of the National Science Foundation. 

This work used the Hive cluster, which is supported by the National Science Foundation under Grant No. 1828187.  This research was supported in part through research cyberinfrastructure resources and services provided by the Partnership for an Advanced Computing Environment (PACE) at the Georgia Institute of Technology, Atlanta, Georgia, USA.

\section{Abbreviations}

\begin{itemize}
    \item EXAFS: extended X-ray absorption fine structure
    \item XANES: X-ray absorption near edge structure
    \item XES: X-ray emissions spectroscopy
    \item XRD: X-ray diffraction
    \item PDF: pair distribution function
    \item DFT: density functional theory
\end{itemize}

\begin{acknowledgement}

The authors thank Dr. Alena Alamgir for her assistance in writing the manuscript.

\end{acknowledgement}




\pagebreak
\bibliography{achemso-demo}

\end{document}